\documentclass[runningheads]{llncs}
\usepackage{amsmath,amssymb,amsfonts}
\usepackage{graphicx}
\usepackage{textcomp}
\usepackage{xcolor}
\usepackage{fancyhdr}
\usepackage[hyphens]{url}

\usepackage{csquotes} 

\usepackage{caption}
\usepackage{subcaption}
\usepackage{algorithm}
\usepackage{algpseudocode}
\usepackage{bm} 
\usepackage{booktabs}
\usepackage{listings}
\usepackage{color}
\usepackage[normalem]{ulem}
\usepackage{tikz}

\definecolor{mygreen}{rgb}{0,0.6,0}
\definecolor{mygray}{rgb}{0.5,0.5,0.5}
\definecolor{mymauve}{rgb}{0.58,0,0.82}

\usepackage{xcolor}
\ifdefined\RELEASEON
    \newcommand\bill[1]{}
    \newcommand\tassos[1]{}
    
\else
    \newcommand\bill[1]{{\color{blue}[Bill]: #1}}
    \newcommand\tassos[1]{{\color{orange}[Tassos]: #1}}
\fi

\usepackage[backref=page]{hyperref}
\hypersetup{
    unicode=false,          
    pdftoolbar=true,        
    pdfmenubar=true,        
    pdffitwindow=true,      
    pdftitle={A Parallel Scan Algorithm in the Tensor Core Unit Model},    
    pdfauthor={Anastasios Zouzias,Bill McColl},     
    pdfsubject={Parallel Algorithms},   
    pdfcreator={Anastasios Zouzias},   
    pdfproducer={Anastasios Zouzias}, 
    pdfkeywords={prefix sum,scan, matrix multiplication,tensor core unit model, tensor cores, systolic arrays}, 
    pdfnewwindow=true,      
    colorlinks=false,       
    linkcolor=red,          
    citecolor=green,        
    filecolor=magenta,      
    urlcolor=blue,          
}

\renewcommand*{\backref}[1]{}
\renewcommand*{\backrefalt}[4]{%
	\ifcase #1 %
		(Not cited) %
	\or
		(Cited on page~#2)%
	\else
		(Cited on pages~#2)%
	\fi
}

\newcommand{\OO}{\mathcal{O}}

\newcommand{\x}{\bm{x}}
\newcommand{\y}{\bm{y}}
\newcommand{\z}{\bm{z}}
\newcommand{\w}{\bm{w}}
\newcommand{\q}{\bm{q}}

\newcommand{\one}{\bm{1}}
\newcommand{\matA}{\bm{A}}
\newcommand{\matL}{\bm{L}}
\newcommand{\matB}{\bm{B}}
\newcommand{\matW}{\bm{W}}
\newcommand{\matT}{\bm{T}}

\usetikzlibrary{calc,shapes.geometric,matrix,positioning,arrows.meta,arrows,shapes.arrows,decorations.pathreplacing,calligraphy,graphs}

\tikzset{
mymat/.style={
  matrix of math nodes,
  text height=2.5ex,
  text depth=0.75ex,
  text width=3.25ex,
  align=center,
  column sep=-\pgflinewidth
  },
mymats/.style={
  mymat,
  nodes={draw,fill=#1}
  }  
}

\begin{document}
\title{A Parallel Scan Algorithm in the Tensor Core Unit Model}
\titlerunning{Parallel Scan in the TCU Model}
%

\author{Anastasios Zouzias\orcidID{0000-0002-3743-3489} \and
William F. McColl
}
\authorrunning{A. Zouzias and W. F. McColl}
%
\institute{Computing Systems Laboratory \\
Zurich Research Center \\
Huawei Technologies, Switzerland \\
\email{\{anastasios.zouzias,bill.mccoll\}@huawei.com}
}
\maketitle              
\begin{abstract}
We present a parallel scan (prefix sum) algorithm in the Tensor Core Unit (TCU) model of computation. The TCU model assumes that multiplication between two square matrices of constant size $s$ is a basic operation. In the $(s^2,\ell )$-TCU model, we show that for inputs of size $n$, the algorithm has depth at most $2\lfloor\log_s(n)\rfloor$ and runs in $\OO(n(1+\ell/s^2) / p + (s^2 + \ell) \log_s (n))$ time assuming $p$ tensor core units. Equivalently, the algorithm performs $\OO(n/s^2)$ multiplications of square matrices of size $s$.

\end{abstract}
%
%
%

%
\section{Introduction}\label{sec:intro}
\begin{figure}[t!]
\begin{subfigure}[t]{0.7\linewidth}
\begin{center}
       \resizebox{!}{0.55\textwidth}{\begin{tikzpicture}[>=latex]
\matrix[mymats=white,anchor=west]
at (0,0) 
(mat1)
{
1 & 2 & 3 & 4 & 5 & 6 & 7 & 8 & 9 & 10 & 11 & 12 & 13 & 14 & 15 & 16\\
};
\matrix[mymats=white,anchor=west]
at (0,-2) 
(mat2)
{
1 & 3 & 6 & 10 & 5 & 11 & 18 & 26 & 9 & 19 & 30 & 42 & 13 & 27 & 42 & 58\\
};
\matrix[mymats=white,anchor=west]
at (0,-4) 
(mat3)
{
1 & 3 & 6 & 10 & 5 & 11 & 18 & 36 & 9 & 19 & 30 & 78 & 13 & 27 & 42 & 136\\
};
\matrix[mymats=white,anchor=west]
at (0,-6) 
(mat4)
{
1 & 3 & 6 & 10 & 15 & 21 & 28 & 36 & 45 & 55 & 66 & 78 & 91 & 105 & 120 & 136\\
};

\node[above=0pt of mat1]
  (cella) {};
\node[above=0pt of mat2]
  (cellb) {};
\begin{scope}[shorten <= -2pt]
\node (elem2-2) [draw, fill =none,below = of mat1-1-2,minimum size=8pt,yshift=1pt, circle,inner sep=-3]{+};
\node (elem2-3) [draw, fill =none,below = of mat1-1-3,minimum size=8pt,yshift=1pt, circle,inner sep=-3]{+};
\node (elem2-4) [draw, fill =none,below = of mat1-1-4,minimum size=8pt,yshift=1pt, circle,inner sep=-3]{+};
\node (elem2-6) [draw, fill =none,below = of mat1-1-6,minimum size=8pt,yshift=1pt, circle,inner sep=-3]{+};
\node (elem2-7) [draw, fill =none,below = of mat1-1-7,minimum size=8pt,yshift=1pt, circle,inner sep=-3]{+};
\node (elem2-8) [draw, fill =none,below = of mat1-1-8,minimum size=8pt,yshift=1pt, circle,inner sep=-3]{+};
\node (elem2-10) [draw, fill =none,below = of mat1-1-10,minimum size=8pt,yshift=1pt, circle,inner sep=-3]{+};
\node (elem2-11) [draw, fill =none,below = of mat1-1-11,minimum size=8pt,yshift=1pt, circle,inner sep=-3]{+};
\node (elem2-12) [draw, fill =none,below = of mat1-1-12,minimum size=8pt,yshift=1pt, circle,inner sep=-3]{+};
\node (elem2-14) [draw, fill =none,below = of mat1-1-14,minimum size=8pt,yshift=1pt, circle,inner sep=-3]{+};
\node (elem2-15) [draw, fill =none,below = of mat1-1-15,minimum size=8pt,yshift=1pt, circle,inner sep=-3]{+};
\node (elem2-16) [draw, fill =none,below = of mat1-1-16,minimum size=8pt,yshift=1pt, circle,inner sep=-3]{+};

\node (elem3-8) [draw, fill =none,below = of mat2-1-8,minimum size=8pt,yshift=1pt, circle,inner sep=-3]{+};
\node (elem3-12) [draw, fill =none,below = of mat2-1-12,minimum size=8pt,yshift=1pt, circle,inner sep=-3]{+};
\node (elem3-16) [draw, fill =none,below = of mat2-1-16,minimum size=8pt,yshift=1pt, circle,inner sep=-3]{+};

\node (elem4-5) [draw, fill =none,below = of mat3-1-5,minimum size=8pt,yshift=1pt, circle,inner sep=-3]{+};
\node (elem4-6) [draw, fill =none,below = of mat3-1-6,minimum size=8pt,yshift=1pt, circle,inner sep=-3]{+};
\node (elem4-7) [draw, fill =none,below = of mat3-1-7,minimum size=8pt,yshift=1pt, circle,inner sep=-3]{+};
\node (elem4-9) [draw, fill =none,below = of mat3-1-9,minimum size=8pt,yshift=1pt, circle,inner sep=-3]{+};
\node (elem4-10) [draw, fill =none,below = of mat3-1-10,minimum size=8pt,yshift=1pt, circle,inner sep=-3]{+};
\node (elem4-11) [draw, fill =none,below = of mat3-1-11,minimum size=8pt,yshift=1pt, circle,inner sep=-3]{+};
\node (elem4-13) [draw, fill =none,below = of mat3-1-13,minimum size=8pt,yshift=1pt, circle,inner sep=-3]{+};
\node (elem4-14) [draw, fill =none,below = of mat3-1-14,minimum size=8pt,yshift=1pt, circle,inner sep=-3]{+};
\node (elem4-15) [draw, fill =none,below = of mat3-1-15,minimum size=8pt,yshift=1pt, circle,inner sep=-3]{+};

\begin{scope}[shorten <= 0.25pt]
    \draw[black] (mat1-1-1.south) -- (elem2-2.north);
    \draw[black] (mat1-1-1.south) -- (elem2-3.north);
    \draw[black] (mat1-1-1.south) -- (elem2-4.north);
    \draw[black] (mat1-1-2.south) -- (elem2-2.north);
    \draw[black] (mat1-1-2.south) -- (elem2-3.north);
    \draw[black] (mat1-1-2.south) -- (elem2-4.north);
    \draw[black] (mat1-1-3.south) -- (elem2-3.north);
    \draw[black] (mat1-1-3.south) -- (elem2-4.north);
    \draw[black] (mat1-1-4.south) -- (elem2-4.north);

    \draw[black] (mat1-1-5.south) -- (elem2-6.north);
    \draw[black] (mat1-1-5.south) -- (elem2-7.north);
    \draw[black] (mat1-1-5.south) -- (elem2-8.north);
    \draw[black] (mat1-1-6.south) -- (elem2-6.north);
    \draw[black] (mat1-1-6.south) -- (elem2-7.north);
    \draw[black] (mat1-1-6.south) -- (elem2-8.north);
    \draw[black] (mat1-1-7.south) -- (elem2-7.north);
    \draw[black] (mat1-1-7.south) -- (elem2-8.north);
    \draw[black] (mat1-1-8.south) -- (elem2-8.north);

    \draw[black] (mat1-1-9.south) -- (elem2-10.north);
    \draw[black] (mat1-1-9.south) -- (elem2-11.north);
    \draw[black] (mat1-1-9.south) -- (elem2-12.north);
    \draw[black] (mat1-1-10.south) -- (elem2-10.north);
    \draw[black] (mat1-1-10.south) -- (elem2-11.north);
    \draw[black] (mat1-1-10.south) -- (elem2-12.north);
    \draw[black] (mat1-1-11.south) -- (elem2-11.north);
    \draw[black] (mat1-1-11.south) -- (elem2-12.north);
    \draw[black] (mat1-1-12.south) -- (elem2-12.north);

    \draw[black] (mat1-1-13.south) -- (elem2-14.north);
    \draw[black] (mat1-1-13.south) -- (elem2-15.north);
    \draw[black] (mat1-1-13.south) -- (elem2-16.north);
    \draw[black] (mat1-1-14.south) -- (elem2-14.north);
    \draw[black] (mat1-1-14.south) -- (elem2-15.north);
    \draw[black] (mat1-1-14.south) -- (elem2-16.north);
    \draw[black] (mat1-1-15.south) -- (elem2-15.north);
    \draw[black] (mat1-1-15.south) -- (elem2-16.north);
    \draw[black] (mat1-1-16.south) -- (elem2-16.north);

    \draw[black] (mat2-1-4.south) -- (elem3-8.north);
    \draw[black] (mat2-1-4.south) -- (elem3-12.north);
    \draw[black] (mat2-1-4.south) -- (elem3-16.north);
    \draw[black] (mat2-1-8.south) -- (elem3-8.north);
    \draw[black] (mat2-1-8.south) -- (elem3-12.north);
    \draw[black] (mat2-1-8.south) -- (elem3-16.north);
    \draw[black] (mat2-1-12.south) -- (elem3-12.north);
    \draw[black] (mat2-1-12.south) -- (elem3-16.north);
    \draw[black] (mat2-1-16.south) -- (elem3-16.north);

    \draw[black] (mat3-1-4.south) -- (elem4-5.north);
    \draw[black] (mat3-1-4.south) -- (elem4-6.north);
    \draw[black] (mat3-1-4.south) -- (elem4-7.north);
    \draw[black] (mat3-1-5.south) -- (elem4-5.north);
    \draw[black] (mat3-1-6.south) -- (elem4-6.north);
    \draw[black] (mat3-1-7.south) -- (elem4-7.north);

    \draw[black] (mat3-1-8.south) -- (elem4-9.north);
    \draw[black] (mat3-1-8.south) -- (elem4-10.north);
    \draw[black] (mat3-1-8.south) -- (elem4-11.north);
    \draw[black] (mat3-1-9.south) -- (elem4-9.north);
    \draw[black] (mat3-1-10.south) -- (elem4-10.north);
    \draw[black] (mat3-1-11.south) -- (elem4-11.north);

    \draw[black] (mat3-1-12.south) -- (elem4-13.north);
    \draw[black] (mat3-1-12.south) -- (elem4-14.north);
    \draw[black] (mat3-1-12.south) -- (elem4-15.north);
    \draw[black] (mat3-1-13.south) -- (elem4-13.north);
    \draw[black] (mat3-1-14.south) -- (elem4-14.north);
    \draw[black] (mat3-1-15.south) -- (elem4-15.north);
\end{scope}
\end{scope}
\path (0,0) node [single arrow,draw,left = of mat1,rotate=270,xshift=3.7cm,minimum height=4cm,rounded corners]{\textbf{Up-sweep}};
\path (0,0) node [single arrow,draw,left = of mat3,rotate=270,xshift=2.5cm,minimum height=2.5cm,rounded corners]{\textbf{Down-sweep}};
\end{tikzpicture}}
     \caption{Input parameter $s=4$.}
     \label{fig:alg_exec_s4_k2}
 \end{center}
\end{subfigure}
\begin{subfigure}[t]{0.9\linewidth}
\centering
\resizebox{\textwidth}{!}{\input{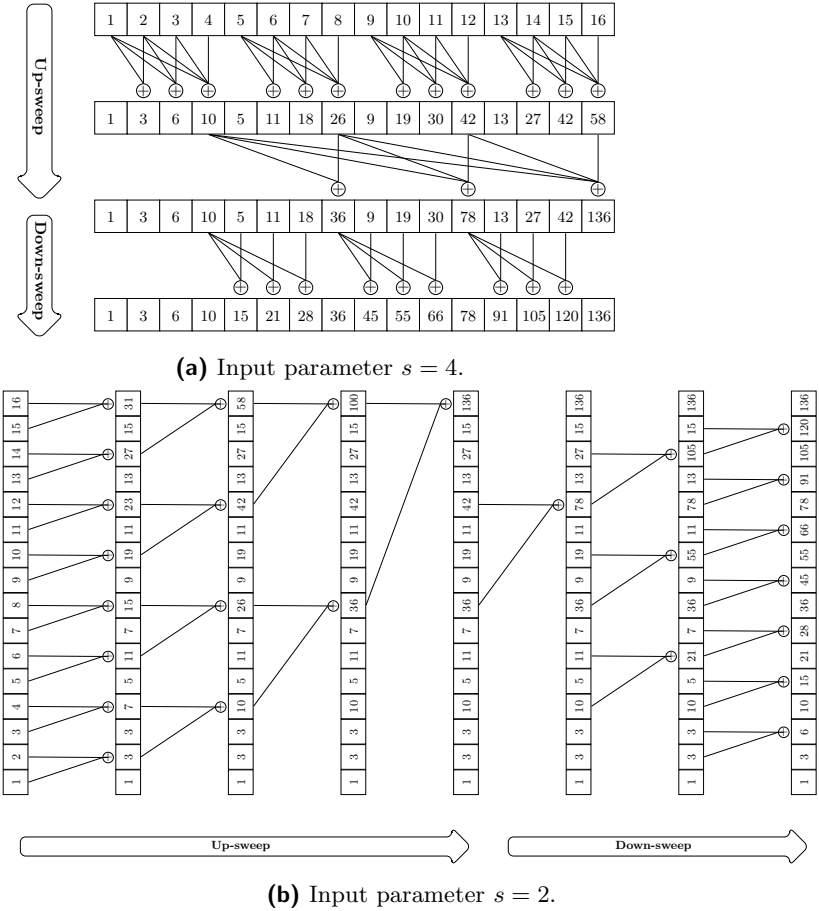}}
\caption{Input parameter $s=2$.}
\label{fig:alg_exec_s2_k4}
\end{subfigure}
\caption{Examples of Algorithm~\ref{alg:tcu_scan} for input $\x=[1,2,\dots , 16]$.}
\label{fig:alg_exec}
\end{figure}

%
Prefix sum (scan) is an important computational primitive in parallel computing with a plethora of applications~\cite{blelloch_prefix_1990,book:gpus}. An extensive literature on parallel scan algorithms provides trade-offs between the depth (length of the critical path of computation) and work (number of binary arithmetic operations) of several approaches in the Parallel Random-Access Machine (PRAM) model of computation. Prefix computation also occurs in carry-lookahead adders where several parallel scan algorithms have been (implicitly) designed (see~\cite{adders:taxonomy:harris03,adders:zimmermann_eth97} and references therein). Moreover, the depth and size trade-offs for parallel optimal prefix circuits are well-understood for binary operations~\cite{scan:snir86,scan:zero_def_circuits06}. In this work, we consider prefix sums in an emerging model of computation. Following the seminal work of~\cite{scan:tcuICS19}, we present a parallel scan algorithm in a recently proposed Tensor Core Unit (TCU) model of computation~\cite{tcu_model:spaa20,tcu_model:europar21}.
The TCU model, denoted\footnote{The first parameter $s^2$ of the TCU model is squared to avoid writing square roots on the matrix sizes.} by $(s^2,\ell )$-TCU, is a standard RAM model where there exists a circuit named tensor unit that performs matrix multiplication between a matrix of size $s\times s$ and $s\times m$ ($m\geq s$) in time $\OO(ms+\ell )$, where $s> 1$ and $\ell \geq 0 $ are two model parameters~\cite{tcu_model:europar21}. The parameter $\ell$ corresponds to the latency of initiating a matrix multiplication operation on the tensor unit. Here, in addition to the runtime analysis of the TCU model, we present a simplified analysis of the work/depth model by assuming that the multiplication of two square matrices of size $s$ is a basic operation and counting the matrix multiplications required by the algorithm. Then, we translate the bounds on the number of matrix multiplications to a time complexity bound of the TCU model.
The reader might wonder why the TCU model is currently an emerging domain-specific model of computation. The primary reason is that deep learning and High-Performance Computing (HPC) workloads have increased the demand for hardware that delivers more efficient matrix multiplication operations~\cite{tpu:v4,alexnet}. Hardware vendors have responded to such demand by manufacturing accelerators with specialized hardware units known as \emph{tensor core units}. A representative list of specialized hardware units includes TPUs~\cite{google_tpus:ISCA17,tpu:v4}, Tensor Cores (TCs)~\cite{nvidia:tesla_v100} and Huawei's Ascend Cube Unit~\cite{huawei_ascend:hpca2021,huawei_ascend:hotchips19} to name a few. In short, today's high-performance hardware accelerators contain tensor core units that allow efficient multiplication of constant-sized square matrices. As advocated recently in~\cite{tcu_model:europar21}, these tensor core units can be employed beyond deep learning and HPC applications to other essential computational primitives (matrix computations, graph algorithms, etc.). Here, we aim to advance this line of work by studying the computational primitive of parallel prefix sums.
The paper's main contribution is the analysis of a parallel scan algorithm (Algorithm~\ref{alg:tcu_scan}) in the TCU model in terms of depth, number of matrix multiplications, work and time complexity. Interestingly enough, Algorithm~\ref{alg:tcu_scan} can be viewed as a generalization of the Brent-Kung scan algorithm~\cite{scan:brent_kung_stoc80}; Brent-Kung is a special case of Algorithm~\ref{alg:tcu_scan} where the matrices have size two, see Figure~\ref{fig:alg_exec} for examples.
Our motivation to study the parallel scan primitive in the TCU model is based on two applications: training gradient boosting trees models and parallel sorting. Indeed, an inspection of the binary tree split computation for training gradient boosting trees reveals that multiple prefix sum operations occur~\cite{xgboost}. For the application of parallel sorting, following Blelloch's reduction of Radixsort to prefix sums~\cite{blelloch_prefix_1990}, we resolve in the affirmative the open question ``can TCU sort?'' raised during the presentation of~\cite{tcu_model:spaa20}.
We conclude this section by introducing our notation. We use the terms \textit{prefix sum} and \textit{scan} interchangeably. By prefix sum, we always refer to \emph{inclusive} prefix sum unless explicitly noted. All results are stated for the addition operator but can be extended to any arbitrary associative operator. Vectors are denoted by lower-case boldface font letters; vectors are always considered column vectors. $\one_s$ denotes the all-ones vector of size $s$. Let $\alpha$ be a scalar, and $\q$ be a vector of size $s-1$; we denote by $[\alpha;\q]$ the column vector of size $s$ whose first entry is $\alpha$ concatenated by $\q$. Matrices are denoted by upper-case bold-face letters. $\matL_s$ is the lower triangular all-ones square matrix of size $s$, including ones on the diagonal. We use zero-based indexing for vectors and matrices. For a vector $\x$, we denote $\x[\text{start}::\text{step}]$ the subvector of $\x$ starting from index \textit{start} with a stride of size \textit{step}. We frequently use the ceiling inequality: $\lceil \alpha \rceil + \lceil \beta \rceil \leq \lceil \alpha + \beta\rceil + 1$ for scalars $\alpha,\beta$.
%
%
%
\section{MatMulScan: Parallel Scan in the TCU Model}\label{sec:tcu_scan}
%
%
\begin{algorithm}[ht]
\begin{algorithmic}[1]
\Procedure{MatMulScan}{$\x$, $s$}
\State Let $n \gets \text{len}(\x)$, $k\gets \log_s(n)$ \Comment{$n:=s^k$}
\For {$t=0,1,.., k-1$}\Comment{$1^{\text{st}}$ Phase (Up-sweep)}
\State $\y \gets $ Gather $\x$ with stride $s^t$ starting from $s^t -1$ \label{lst:line:1st_gather}
\State $\z \gets \Call{BatchMatMul}{\y, \matL_s}$\label{lst:line:batch_MM_up_sweep}
\State Scatter $\z$ into $\x$ with stride $s^t$ starting from $s^t-1$
\EndFor
\For {$t=k-1, \dots ,2 ,1$}\Comment{$2^{\text{nd}}$ Phase (Down-sweep)}
\State $\y \gets $ Gather $\x$ with stride $s^{t-1}$ starting from $s^t -1$ \label{lst:line:2nd_gather}
\State $\z \gets \Call{BatchMatMul}{\y, \matB_s}$\label{lst:line:batch_MM_down_sweep}
\State Scatter $\z$ into $\x$ with stride $s^{t-1}$ starting from $s^t-1$ 
\EndFor
\State \textbf{Output:} Return $\x$
\EndProcedure
%
%
\Procedure{BatchMatMul}{$\y, \matA_s$}\Comment{$s\times s$ matrix $\matA_s$}
\State Let $m\gets \text{len}(\y)$, $s\gets \text{numCols}(\matA_s)$
\State Zero-pad $\y$ to size $s^2\lceil m/s^2\rceil$\label{lst:line:zero_pad} \Comment{Or, zero-pad $\y$ to size $s\lceil m/s\rceil$}
\State $\matT\gets $ View $\y$ as a $(\lceil m/s^2\rceil, s, s)$-tensor \Comment{Or, view $\y$ as $s\times \lceil m/s\rceil$ matrix}\label{lst:line:reshape_tensor}
\State $\matW \gets $ Batch matrix multiplication $\matA_s$ and $\matT$ \label{lst:line:batch_matmul}
\State $\z \gets$ Flatten $\matW$ to $m$-vector  (drop zero-padding)\label{lst:line:flatten_tensor}
\State \textbf{Output:} Return $\z$
\EndProcedure
\end{algorithmic}
\caption{Parallel Matrix Multiplication Scan}\label{alg:tcu_scan}
\end{algorithm}
%
In this section, we present a parallel scan algorithm (Algorithm~\ref{alg:tcu_scan}) designed to take advantage of the computational speedup offered by the matrix multiplication circuit of the TCU model. All numerical operations of Algorithm~\ref{alg:tcu_scan} are multiplications between two square matrices of size $s$. Surprisingly enough, only two special (constant) matrices take place as the left operand in all matrix multiplications. These two special matrices encode the computation of local prefix sums and a scalar/vector add operation.
Let's first define the matrix that encodes the (local) prefix sum operator. Given a vector $\w$ of size $s$, it is straightforward to verify that the prefix sum of $\w$ equals the matrix product $\matL_s \w$ (recall $\matL_s$ is the lower triangular all-ones square matrix). Next, we encode the addition between a vector $\q$ of size $s-1$ and a scalar $\alpha$, i.e., $\q + \alpha \one_{s-1}$ as follows. The scalar/vector addition of $\alpha$ and $\q$ can be extracted from the result of the matrix-vector product $\matB_{s} [\alpha; \q]$ where $\matB_s$ is a square matrix of size $s$ defined as:
%
\begin{equation*}\label{eq:bcast_mat}
\small
\matB_s:=\begin{bmatrix}
1 & 0 & 0 &\dots & 0\\
1 & 1 & 0 &  \dots & 0 \\
1 & 0 & 1 &  \dots & 0 \\
\vdots & 0 & 0 &\ddots& 0 & \\
1 & 0 & \dots & 0 & 1 
\end{bmatrix}.
\end{equation*}
%
Now we are ready to define the main algorithm (\textsc{MatMulScan}). Algorithm~\ref{alg:tcu_scan} consists of two phases, as is typical in work-efficient parallel scan algorithms: the up-sweep phase (Lines 3-7) and the down-sweep (Lines 8-11) phase. In the first up-sweep phase, the prefix sums of the indices with exponentially increasing sizes are computed: $s,s^2,s^3,\dots $, etc. At the end of the first phase, the prefix sums are correct only on an exponentially increasing set of indices. The remaining indices contain a ``local prefix sum'', i.e., a prefix sum of $s$ contiguous indices that will be corrected in the second phase. The second down-sweep phase broadcasts and adds all the precedent prefix sums to the remaining local prefix sums. At each round of both phases, a strided subset of the input vector is viewed as a matrix/tensor and pre-multiplied with a constant matrix of size $s$ as is described in the procedure \textsc{BatchMatMul} (Lines 15-22).
%

%
%
\begin{figure}[h!]
\begin{center}
  \includegraphics[width=0.9\linewidth]{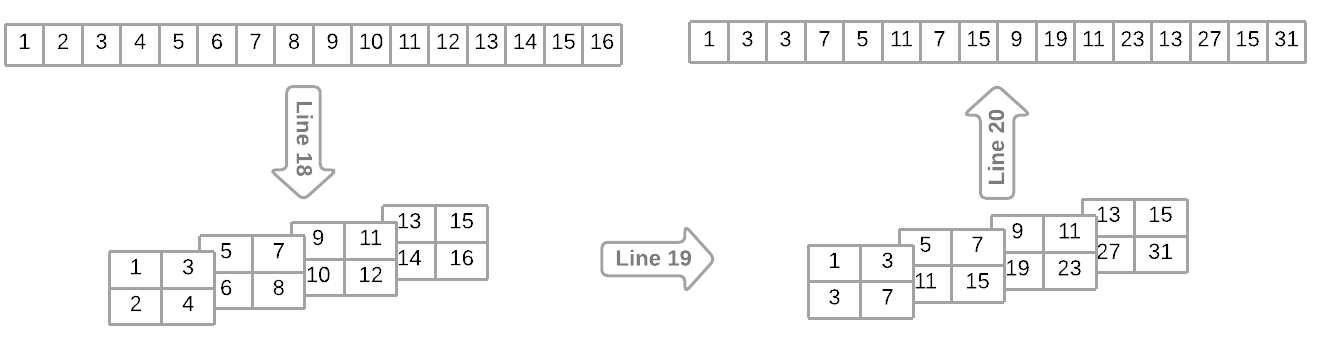}
\end{center}
\caption{Execution of \textsc{BatchMatMul$(\y=[1;2;\dots ;16], \matA_2=\matL_2)$}.}
\label{fig:tensor_transform}
\end{figure}
%
The \textsc{BatchMatMul} procedure takes a vector $\y$ and a square matrix $\matA_s$ of size $s$ as input. \textsc{BatchMatMul} performs a multiplication between $\matA_s$ and a reshaped tensor view of $\y$. The vector/tensor reshaping operations of the \textsc{BatchMatMul} method (Lines~\ref{lst:line:reshape_tensor} and \ref{lst:line:flatten_tensor}) need more clarification. In Line~\ref{lst:line:reshape_tensor}, a vector is viewed as a three-dimensional tensor as follows. The zero-padded vector $\y$ (Line~\ref{lst:line:zero_pad}) is split into multiple chunks of size $s^2$, and each chunk is viewed as an $s\times s$ matrix in column-major layout. Each $s\times s$ matrix is stacked into a three-dimensional tensor of size $(\lceil m/s^2\rceil ,s,s)$ following the natural ordering, i.e., the first chunk is assigned to index zero of the first dimension of the tensor, the second chunk to index one, etc.
Figure~\ref{fig:tensor_transform} provides an illustrative example of the execution of \textsc{BatchMatMul} with inputs: a vector of size $n=16$ and $\matA_2=\matL_2$. In addition, we provide an end-to-end functional implementation of Algorithm~\ref{alg:tcu_scan} as a reference implementation in Appendix~\ref{appendix:tcu_scan_numpy}. 
%
\subsection{Analysis}\label{sec:analysis}
%
In this section, we analyse Algorithm~\ref{alg:tcu_scan} in terms of depth, the number of matrix multiplications required, work and time complexity in the TCU model. In the analysis, we ignore\footnote{That said, the cost of memory operations (memory coalescing, bank conflicts, etc.) is crucial to achieving high performance in an actual implementation.} the cost of the gather-scatter memory operations and the cost of vector/tensor reshaping operations. Recall that multiplication between two square matrices of size $s$ is a basic operation.
%
\begin{lemma}\label{lem:tcu_scan}
Fix an integer $s\geq 2$. Let $\x$ be a vector of size $n=s^k$ for some $k$. Algorithm~\ref{alg:tcu_scan} has depth $2k-1$ in the TCU model and performs at most $\lceil \frac{2n}{s(s-1)}\rceil +2k-2$ matrix multiplications. Moreover, the number of scalar binary additions executed by Algorithm~\ref{alg:tcu_scan} is $\lceil n(1+s/2) \rceil + \OO(s^3 \log_s (n))$.
\end{lemma}
%
%
\begin{proof}
%
The first phase takes $k$ steps, and the second takes $k-1$ steps. In total, the depth is $2k-1=2\log_s(n)-1=2\log_2(n)/\log_2(s)-1$ in the TCU model.
Let's calculate the number of matrix multiplications required per phase. In the first phase and at the $t$-th iteration (Line~\ref{lst:line:1st_gather}) $\y$ has length $\lfloor (n-(s^t-1))/s^t\rfloor$. Hence, at most $\lceil n/s^{t+2}\rceil $ matrix multiplications occur in Line~\ref{lst:line:batch_MM_up_sweep}. In total, the first phase requires at most $\lceil \frac{n}{s^2}\sum_{t=0}^{k-1}\frac1{s^t}\rceil \leq \lceil\frac{n}{s(s-1)}\rceil + k -1$ matrix multiplications by multiple applications of the ceiling inequality. Similarly, in the second phase and at the $t$-th iteration (Line~\ref{lst:line:2nd_gather}), $\y$ has length $\lfloor (n-(s^t-1))/s^{t-1}\rfloor $. Hence, at most $\lceil n/s^{t+1}\rceil $ matrix multiplications occur in Line~\ref{lst:line:batch_MM_down_sweep}. In total, the second phase requires at most $\lceil \frac{n}{s^2}\sum_{t=1}^{k-1}\frac1{s^{t-1}}\rceil\leq \lceil \frac{n}{s(s-1)} \rceil+ k-2$ by using the ceiling inequality. In total, at most $\lceil \frac{2n}{s(s-1)}\rceil + 2k-2$ matrix multiplications are required by the algorithm.
Now, let's compute the work of the algorithm in terms of scalar binary additions when simulated in the RAM model, i.e., bound the number of arithmetic operations of the matrix multiplications. The number of scalar binary additions of matrix-vector multiplication between $\matL_s$ and a vector of size $s$ takes $s(s-1)/2$ scalar additions. Therefore, the work of the first phase is
%
\begin{equation*}
    \left(\lceil\frac{n}{s(s-1)}\rceil + k -1\right) \cdot s \cdot \frac{s(s-1)}{2} = \lceil ns/2 \rceil + \OO(ks^3).
\end{equation*}
%
Similarly, the work of the second phase is
%
\begin{equation*}
    \left(\lceil\frac{n}{s(s-1)}\rceil + k -2\right) \cdot s \cdot (s-1) = n + \OO(ks^2),
\end{equation*}
%
since each matrix multiplication between $\matB_s$ and a square matrix of size $s$ takes $s(s-1)$ scalar additions. In total, the work of the algorithm is $\lceil n(1+s/2)\rceil + \OO(s^3\log_s(n))$.
\end{proof}
%
We defer the correctness proof of Algorithm~\ref{alg:tcu_scan} to Appendix~\ref{appendix:correctness}.
Next, we translate the analysis of Lemma~\ref{lem:tcu_scan} into a time complexity bound in the $(s^2,\ell )$-TCU model with a minor modification of Algorithm~\ref{alg:tcu_scan}. We view the tensor $\matT$ (Line~$18$) into a single rectangular matrix of size $s\times \lceil m/s \rceil$ by stacking over its first dimension in Line~\ref{lst:line:batch_matmul}. The stacking allows us to avoid excessive matrix multiplication invocations, i.e., increased latency cost.
%
\begin{theorem}\label{thm:tcu_time}
Fix an integer $s\geq 2$. Let $\x$ be a vector of size $n=s^k$ for some $k$. Algorithm~\ref{alg:tcu_scan} takes $\OO(n + \ell k)$ time in the $(s^2,\ell )$-TCU model.
\end{theorem}
%
\begin{proof}
Let's bound the latency cost and the matrix multiplication cost separately. Recall that the depth of the computation is $2k-1$. At each round, the batched matrix multiplication (Line~\ref{lst:line:batch_matmul}) can be viewed as a multiplication invocation between an $s\times s$ and $s\times \lceil m/s \rceil $ matrix ($m$ is defined in Line~\ref{lst:line:zero_pad}). Hence, the latency cost is $(2k-1)\ell$ since $2k-1$ matrix multiplication invocations take place. Next, let's bound the time cost of matrix multiplications. For all matrix multiplications of the first phase, the time required is
$\sum_{j=0}^{k-1} \lceil \frac{n}{s^{j}}\frac1{s}\rceil s   \leq  s\lceil \frac{n}{s} \sum_{j=0}^{k-1} \frac1{s^{j}}  \rceil  + sk - s = \OO(n)$, where the first inequality follows by the ceiling inequality and the second inequality since $\sum_{j=0}^{k-1} \frac1{s^{j}}\leq 2$ provided that $s\geq 2$. Similarly, the time required for the matrix multiplications in the second phase is $\sum_{j=0}^{k-2} \lceil \frac{n}{s^{j}}\frac1{s}\rceil s=\OO(n)$. In total, the time complexity of Algorithm~\ref{alg:tcu_scan} in the $(s^2,\ell )$-TCU model is $\OO(n+\ell \log_s(n))$.
\end{proof}
%
%
\subsection{Extend Algorithm~\ref{alg:tcu_scan} to Arbitrary Input Length}\label{sec:extension}
%
%
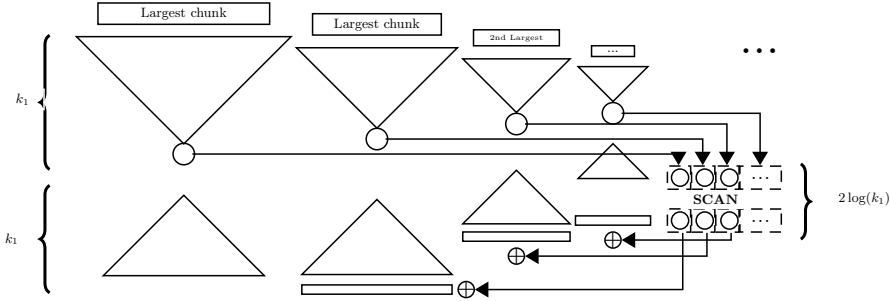
\begin{figure}[ht]
\begin{center}
  \resizebox{\linewidth}{!}{\begin{tikzpicture}
\node[isosceles triangle,
    isosceles triangle apex angle=90,
    draw,
    rotate=270,
    fill=none,
    minimum size =1cm] (T1) at (0,0) {};
\node (circ1-1) [draw, shape = circle, fill = none, minimum size = 0.2cm, inner sep=0pt,below = of T1] at ([yshift=0.4cm]T1){};
\node (rec1-1) [draw,shape = rectangle,above = of T1,scale=0.4,minimum width = 4cm] at ([yshift=-0.5cm]T1){Largest chunk}; 
\node[isosceles triangle,
    isosceles triangle apex angle=90,
    draw,
    rotate=90,
    fill=none,
    minimum size =0.75cm] (T5) at ([yshift=-0.85cm]circ1-1) {};
\node[isosceles triangle,
    isosceles triangle apex angle=90,
    draw,
    rotate=270,
    fill=none,
    minimum size =0.75cm] (T2) at (1.8,0) {};
\node (circ1-2) [draw, shape = circle, fill = none, minimum size = 0.2cm, inner sep=0pt,below = of T2] at ([yshift=0.54cm]T2){};
\node (rec1-2) [draw,shape = rectangle,above = of T2,scale=0.4,minimum width = 3cm] at ([yshift=-0.6cm]T2){Largest chunk}; 
\node[isosceles triangle,
    isosceles triangle apex angle=90,
    draw,
    rotate=90,
    fill=none,
    minimum size =.7cm] (T6) at ([yshift=-1cm]circ1-2) {};
\node (rec1-5) [draw,shape = rectangle,above = of T2,scale=0.4,minimum width = 3.5cm] at ([yshift=-1.45cm]T6){};
\node (circ1-5) [draw, shape = circle, fill = none, minimum size = 0.15cm, inner sep=-3,right = of rec1-5] at ([xshift=-0.25cm]rec1-5){\tiny{+}};
\node[isosceles triangle,
    isosceles triangle apex angle=90,
    draw,
    rotate=270,
    fill=none,
    minimum size =0.5cm] (T3) at (3.1,0) {};
\node (circ1-3) [draw, shape = circle, fill = none, minimum size = 0.2cm, inner sep=0pt,below = of T3] at ([yshift=0.68cm]T3){};
\node (rec1-3) [draw,shape = rectangle,above = of T2,scale=0.4,minimum width = 2cm] at ([yshift=-0.7cm]T3){\tiny{2nd Largest}}; 
\node[isosceles triangle,
    isosceles triangle apex angle=90,
    draw,
    rotate=90,
    fill=none,
    minimum size =.5cm] (T7) at ([yshift=-.75cm]circ1-3) {};
\node (rec1-6) [draw,shape = rectangle,scale=0.4,minimum width = 2.5cm] at ([yshift=-.3cm]T7){};
\node (circ1-6) [draw, shape = circle, fill = none, minimum size = 0.15cm, inner sep=-3,below = of rec1-6] at ([yshift=0.9cm]rec1-6){\tiny{+}};
\node[isosceles triangle,
    isosceles triangle apex angle=90,
    draw,
    rotate=270,
    fill=none,
    minimum size =0.25cm] (T4) at (4,0) {};
\node (circ1-4) [draw, shape = circle, fill = none, minimum size = 0.2cm, inner sep=0pt,below = of T4] at ([yshift=0.78cm]T4){};
\node (rec1-4) [draw,shape = rectangle,above = of T2,scale=0.4,minimum width = 1cm] at ([yshift=-0.8cm]T4){$...$};
\node[isosceles triangle,
    isosceles triangle apex angle=90,
    draw,
    rotate=90,
    fill=none,
    minimum size =.25cm] (T8) at ([yshift=-.5cm]circ1-4) {};
\node (rec1-7) [draw,shape = rectangle,scale=0.4,minimum width = 1.75cm] at ([yshift=-.50cm]T8){};
\node (circ1-7) [draw, shape = circle, fill = none, minimum size = 0.15cm, inner sep=-3,below = of rec1-7] at ([yshift=0.9cm]rec1-7){\tiny{+}};

\node (s1) [draw,shape = rectangle,densely dashed,right = of rec1-7] at ([xshift=-0.5cm]rec1-7){};
\draw[black] ([xshift=0.625cm]rec1-7)  circle(0.08cm);
\node (s2) [draw,shape = rectangle,densely dashed,right = of s1] at ([xshift=-0.89cm]s1){};
\draw[black] ([xshift=0.855cm]rec1-7)  circle(0.08cm);
\node (s3) [draw,shape = rectangle,densely dashed,right = of s2] at ([xshift=-0.89cm]s2){};
\draw[black] ([xshift=1.085cm]rec1-7)  circle(0.08cm);
\node (dots1) [draw,shape = rectangle,densely dashed,right = of s3,scale=0.9] at ([xshift=-0.89cm]s3){\tiny{...}};

\node (Text1) [draw = white,shape = rectangle,right = of rec1-7,scale=0.4,minimum width = 1cm] at ([yshift=0.2cm,xshift=-0.3cm]rec1-7) {\textbf{SCAN}};

\node (s4) [draw,shape = rectangle,densely dashed,right = of rec1-7] at ([yshift=0.4cm,xshift=-0.5cm]rec1-7){};
\draw[black] ([yshift=0.4cm,xshift=0.625cm]rec1-7)  circle(0.08cm);
\node (s5) [draw,shape = rectangle,densely dashed,right = of s1] at ([yshift=0.4cm,xshift=-0.89cm]s1){};
\draw[black] ([yshift=0.4cm,xshift=0.855cm]rec1-7)  circle(0.08cm);
\node (s6) [draw,shape = rectangle,densely dashed,right = of s2] at ([yshift=0.4cm,xshift=-0.89cm]s2){};
\draw[black] ([yshift=0.4cm,xshift=1.085cm]rec1-7)  circle(0.08cm);
\node (dots2) [draw,shape = rectangle,densely dashed,right = of s6,scale=0.9] at ([xshift=-0.89cm]s6){\tiny{...}};

\node (dots1) [draw=white,shape = rectangle,right = of rec1-4,scale=0.9] at ([xshift=0.075cm]rec1-4){\Large{...}};
\node[anchor=east] at ([xshift=0.08cm]circ1-1) (A1) {};
  \node[anchor=west] at ([xshift=-0.12cm]s4) (A2) {};
  \draw[-Triangle] (A1) -| (A2);

\node[anchor=east] at ([xshift=0.08cm]circ1-2) (A3) {};
  \node[anchor=west] at ([xshift=-0.12cm]s5) (A4) {};
  \draw[-Triangle] (A3) -| (A4);

\node[anchor=east] at ([xshift=0.08cm]circ1-3) (A5) {};
  \node[anchor=west] at ([xshift=-0.12cm]s6) (A6) {};
  \draw[-Triangle] (A5) -| (A6);

\node[anchor=east] at ([xshift=0.08cm]circ1-4) (A7) {};
  \node[anchor=west] at ([xshift=-0.12cm]dots2) (A8) {};
  \draw[-Triangle] (A7) -| (A8);

\node[anchor=east] at ([xshift=0.15cm]s1) (A9) {};
  \node[anchor=west] at ([xshift=-0.15cm]circ1-5) (A10) {};
  \draw[Triangle-] (A10) -| (A9);

\node[anchor=east] at ([xshift=0.15cm]s2) (A11) {};
  \node[anchor=west] at ([xshift=-0.15cm]circ1-6) (A12) {};
  \draw[Triangle-] (A12) -| (A11);

\node[anchor=east] at ([xshift=0.15cm]s3) (A13) {};
  \node[anchor=west] at ([xshift=-0.15cm]circ1-7) (A14) {};
  \draw[Triangle-] (A14) -| (A13);

\draw [decorate,thick,
    decoration = {brace}] (-1.25,-2) --  (-1.25,-1);
\draw [decorate,thick,
    decoration = {brace}] (-1.25,-0.85) --  (-1.25,0.4);
\draw [decorate,thick,
    decoration = {brace}] (5.75,-0.8) --  (5.75,-1.5);

\node (Text2) [draw = white,shape = rectangle,right = of rec1-7,scale=0.4,minimum width = 1cm] at ([xshift=0.1cm]Text1) {$2\log(k_1)$};

\node (Text3) [draw = white,shape = rectangle,left = of rec1-1,scale=0.4,minimum width = 1cm] at ([yshift=-0.8cm,xshift=-0.3cm]rec1-1) {$k_1$};

\node (Text4) [draw = white,shape = rectangle,left = of rec1-5,scale=0.4,minimum width = 1cm] at ([yshift=0.475cm,xshift=-2.2cm]rec1-5) {$k_1$};
\end{tikzpicture}}
\end{center}
\caption{Execution diagram of the general case leveraging Algorithm~\ref{alg:tcu_scan} as a building block. The diagram demonstrates that after the up-sweep phase of the first largest chunks of size $s^{k_1}$, the prefix sum computation of the maximum values of each chunk (excluding the first one) can be interleaved with the down-sweep computation of the largest chunks.}
\label{fig:tcu_scan_general}
\end{figure}
%
In this section, we extend Algorithm~\ref{alg:tcu_scan} for arbitrary input sizes (non-powers of $s$). The approach is based on a folklore approach, see for example~\cite[Chapter~11]{book:gpus}. Let $n$ be an arbitrary positive number. Write $n$ in base $s$ as $n= \sum_{i} \mu_i s^{k_i}$ where $k_1:= \lfloor \log_s(n)\rfloor$, $0\leq \mu_i< s$, and $k_1>k_2>\dots \geq 0$. We assume that $n$ is given in base $s$.
The algorithm is depicted in Figure~\ref{fig:tcu_scan_general} and consists of the following four steps:
%
\begin{enumerate}
\item 
    Execute Algorithm~\ref{alg:tcu_scan} in parallel for each contiguous segment of sizes: $\mu_1$ times on chunks of size $s^{k_1}$, $\mu_2 $ times on chunks of size $s^{k_2},\dots$ etc.
\item
    Gather the maximum values of each segment after the corresponding $1^{\text{st}}$ phase of Algorithm~\ref{alg:tcu_scan} into a vector $\w$ of size at most $k_1(s-1)$. Indeed, there are at most $s-1$ multiples on each segment size and at most $k_1$ distinct segment sizes.
 \item 
    Zero-pad the vector of the maximum values to the smallest integer $q$ so that $k_1(s-1) \leq s^q$ holds. Run Algorithm~\ref{alg:tcu_scan} with the zero-padded input vector of length $s^q$, drop the zero-padding and write back the results on $\w$.
  \item
    For each $i$-th entry of $\w$, in parallel, broadcast and add the value $w_i$ on the $i+1$ chunk of size $s^{k_{i+1}}$ using the BatchMatMul procedure of Algorithm~\ref{alg:tcu_scan}.
\end{enumerate}
%
Let us now analyse the above algorithm in terms of depth, number of matrix multiplications and runtime in the TCU model.

%
\paragraph{Depth.}
%
The first step has depth at most $2k_1-1$ since the largest chunks have size $s^{k_1}$ (Lemma~\ref{lem:tcu_scan}). The execution of the second and third steps can be overlapped with the $2^{\text{nd}}$ phase of the execution of the first step for large enough $n$. The fourth step takes an extra round to perform a scalar/vector addition using matrix multiplications. In total, the depth is $2k_1-1+1=2\lfloor \log_2(n)/\log_2(s) \rfloor$.
%
\paragraph{Matrix Multiplications.}
%
Next, we upper bound the number of matrix multiplications. A direct application of Lemma~\ref{lem:tcu_scan} on the multiple segments of size $s^{k_1},s^{k_2}, \dots$ implies that the number of matrix multiplications is at most
%
\begin{eqnarray*}
 \sum_{i\geq 1} \mu_i \left( \lceil \frac{2s^{k_i}}{s(s-1)}\rceil + 2k_i -1\right) & \leq & \left\lceil\sum_{i\geq 1} \mu_i  \frac{2s^{k_i}}{s(s-1)}\right\rceil  +2\sum_{i\geq 1} \mu_i k_i   \\
  & \leq &  \left\lceil\sum_{i\geq 1} \mu_i  \frac{2s^{k_i}}{s(s-1)}\right\rceil +  2s\sum_{i\geq 1} k_i  \\
 & \leq & \lceil \frac{2n}{s(s-1)}\rceil + \OO(s\log_s^2(n)) 
\end{eqnarray*}
%
where the first inequality follows from the ceiling inequality; the second inequality uses the fact that $\mu_i\leq s$; the third inequality follows since $k_i\leq k_1$, and there are at most $k_1$ terms in the sum. The number of matrix multiplications is negligible on steps $2$ and $3$ since the input size is $\OO(sk_1)$. The fourth step performs a scalar/vector addition with matrix multiplications. Hence it takes at most
%
\begin{eqnarray*}
 \sum_{i\geq 1} \mu_i  \left\lceil \frac{s^{k_i}}{(s-1)^2}\right\rceil 
 & \leq & \left\lceil\frac{2}{s(s-1)}\sum_{i\geq 1} \mu_i  s^{k_i}\right\rceil  + \sum_{i\geq 1} \mu_i
 = \lceil \frac{2n}{s(s-1)}\rceil + \OO(s\log_s(n)),
\end{eqnarray*}
%
where in the first inequality, we used the ceiling inequality and the fact that $\frac1{(s-1)^2}\leq \frac{2}{s(s-1)}$ for $s\geq 2$. In total, the number of matrix multiplications is $\lceil \frac{4n}{s(s-1)}\rceil + \OO(s\log_s^2(n))$.
%
%
\paragraph{Time Analysis in TCU Model.}
%
Apply Theorem~\ref{thm:tcu_time} on the $\mu_i$ segments of size $s^{k_i}$ implies that the time complexity is at most in the order of $\sum_{i} \mu_i \left(s^{k_i} + \ell k_i\right) = n + \ell \sum_{i}\mu_i k_i\leq n+\ell s\sum_{i} k_i\leq n+\ell s k_1^2$, where the first inequality holds since $\mu_i< s$ and the second inequality since $k_i\leq k_1$ and there are at most $k_1$ terms in the sum. In total, Step $1$ takes $\OO(n+\ell s \log_s^2(n))$ time.
Steps $2$ and $3$ are low-order terms and require $\OO(s\log_s(n) + \ell \log_s\log_s(n))$ time. Next, we bound step $4$. At each segment of size $s^{k_i}$, we view each segment as an $(s-1)\times \lceil s^{k_i}/ (s-1)\rceil$ column-major matrix. Then, we prepend the constant row to this matrix that contains the broadcasted value of the previous segment, resulting in an $s\times \lceil s^{k_i}/ (s-1)\rceil$ matrix. Similarly, the running time of step $4$ in the TCU model is
%
\begin{eqnarray*}
    \sum_{i>1} \mu_i  \left( \lceil \frac{s^{k_i}}{(s-1)}\rceil +\ell\right) \leq \sum_{i\geq 1} \mu_i  \lceil \frac{s^{k_i}}{(s-1)}\rceil +\ell\sum_{i\geq 1} \mu_i = \OO(n/s+\ell s \log_s(n)).
\end{eqnarray*}
%
The above discussion is summarized in the following corollary.
%
\begin{corollary}\label{cor:single_tcu}
Fix an integer $s\geq 2$. Let $\x$ be a vector of size $n$. There is an algorithm in the $(s^2,\ell )$-TCU model that has depth at most $2\lfloor \log_s(n)\rfloor$, and takes $\OO(n + s\ell  \log_s^2(n))$ time. Equivalently, the algorithm performs $\OO(n/s^2)$ matrix multiplications.
\end{corollary}
%
Corollary~\ref{cor:single_tcu} reveals that a single tensor core unit does not provide sufficient parallelism for prefix sum computation since the runtime has an additional $\OO(s\ell \log_s^2(n))$ penalty. Therefore, an extension of the TCU model that consists of $p$ parallel tensor core units is essential to provide sufficient parallelism. Recall that the depth of the computation is at most $2\lfloor \log_s(n)\rfloor$ and $\OO(n/s^2)$ matrix multiplications are required in total. Each matrix multiplication takes time $\OO(s^2 + \ell)$. Hence an application of Brent's theorem~\cite{brent_thm74} implies that the runtime is $\OO(n(1 + \ell / s^2)/p + (s^2 + \ell) \log_s (n))$ when having $p$ parallel $(s^2,\ell)$-TCUs as advertised in the abstract.
%
\subsection{Discussion}\label{sec:discussion}
%
We shortly discuss some practical considerations of Algorithm~\ref{alg:tcu_scan}. First, Algorithm~\ref{alg:tcu_scan} on the case where $s=2$ corresponds to the Brent-Kung scan algorithm~\cite{adders:brent_kung82}. Moreover, for inputs of size $n$ that is a power of $s$, the fan-in\footnote{Fan-in is the maximum number of inputs an adder can have. Similarly, fan-out is the maximum number of outputs.} (and fan-out) on all the computations of Algorithm~\ref{alg:tcu_scan} viewed as a circuit with adder nodes are upper-bounded by $s$, see Figure~\ref{fig:alg_exec_s4_k2}. In a nutshell, there is a trade-off between the fan-in/-out and the depth of the computation in Algorithm~\ref{alg:tcu_scan}. This trade-off is explicit in Lemma~\ref{lem:tcu_scan}.
Next, we briefly discuss several implementation issues that could arise in an efficient implementation of Algorithm~\ref{alg:tcu_scan}. Developing a high-performant implementation of parallel scan is a highly non-trivial task and requires a deep understanding of the underlying hardware~\cite{book:gpus}. As is evident by the definition of the matrices $\matL_s$ and $\matB_s$, the utilization of the tensor core unit is low due to the sparsity structure of these matrices. In the first phase, the tensor core unit could be at most $50\%$ utilized due to the lower triangular operand $\matL_s$. The utilization is extremely low in the second phase, roughly speaking $\OO(1/s)$, since the tensor core unit is used for a scalar/vector add operation. However, in a practical implementation, the second phase's scalar/vector add operation can typically be efficiently performed using a vector unit if one exists in proximity to the tensor core unit. Last but not least, the scatter/gather memory operations of Algorithm~\ref{alg:tcu_scan} could be a critical bottleneck to achieving high performance if the tensor core units do not efficiently implement these operations.
%

%
\section{Related work}
%
%
\begin{table}[ht]
\centering
\begin{tabular}{|c|c|c|c|}
\toprule
\textbf{Method}          & \textbf{Depth$^*$} & \textbf{Work} & \textbf{Comments} \\ \midrule
{\small Sklansky~\cite{adders:sklansky60} }     & $\log_2(n)$ & $n\log_2(n)/2$ &  Unbounded fan-out \\ \hline \hline
{\small  Hillis-Steele~\cite{scan:hillis_steele86}}       & $\log_2(n)$ & $n\log_2(n) - n + 1$ &  a.k.a. Kogge-Stone~\cite{adders:kogge_stone73} \\ \hline \hline
Blelloch~\cite{blelloch_prefix_1990}       & $2\log_2(n)$ & $2(n-1)$ &  Exclusive Scan \\ \hline \hline
Brent-Kung~\cite{adders:brent_kung82,scan:brent_kung_stoc80}       & $2\log_2(n)-1$ & $2 n-\log_2(n)-2$ &  Inclusive Scan \\ \hline \hline
\hline
\cite[Alg.~7]{scan:tcuICS19}    & $5 \lceil n/ s^3 \rceil $ & $\OO (ns)$ & TCU / GEMMs \\ \hline \hline
Alg.~\ref{alg:tcu_scan} ($n=2^k$)    & $2\log_2(n) - 1$ & $2n + \OO(\log_2(n))$ & TCU / (Fig.~\ref{fig:alg_exec_s2_k4})
\\ \hline \hline
Alg.~\ref{alg:tcu_scan} ($n=4^k$)    & $\log_2(n) - 1$ & $3n + \OO(\log_2(n))$ & TCU / $s=4$ (Fig.~\ref{fig:alg_exec_s4_k2})
\\ \hline \hline
Alg.~\ref{alg:tcu_scan} ($n=s^k$)    & $2\log_s(n) - 1$& ${\small n(1+s/2)+\OO(s^3\log_s(n))}$ &  TCU / Lemma~\ref{lem:tcu_scan}\\ \hline
\bottomrule
\end{tabular}
\caption{Comparison of a few representative parallel prefix sum algorithms. Work is measured in terms of binary additions. (*) Depth is measured: in the PRAM model for prior work; and in the TCU model for Algorithm~\ref{alg:tcu_scan} and \cite[Alg.~7]{scan:tcuICS19}.} \label{tab:algo_comparison}
\end{table}
%

The study of accelerating the prefix sum (and reduction) operations using the tensor cores was first initiated in the seminal paper of~\cite{scan:tcuICS19} (to the best of our knowledge). The authors of~\cite{scan:tcuICS19} designed scan algorithms for the GPUs architecture, i.e., they proposed\footnote{The main goal of the authors is to provide highly optimized kernels and, hence, use the terms of warp/block/grid of the CUDA programming model.} a warp-level scan algorithm~\cite[Algorithm~6]{scan:tcuICS19}, and a block-level scan algorithm~\cite[Algorithm~7]{scan:tcuICS19}. Moreover, they briefly mentioned that the device/grid level algorithm is based on the textbook approach, see Section~\ref{sec:extension}. Here, we compare Algorithm~\ref{alg:tcu_scan} against Algorithm 7 of~\cite{scan:tcuICS19} in asymptotic analysis. A minor difference with our work is that GEneral Matrix Multiplication (GEMM) is considered as a basic operation in~\cite{scan:tcuICS19}. GEMM is, roughly speaking, a matrix multiplication followed by matrix addition on the output matrix product. Indeed, most tensor core units offer an efficient matrix multiplication and accumulation of the output result with no additional performance cost.
We should highlight that comparing the current work and~\cite{scan:tcuICS19} is not straightforward. The reason is that the goal of~\cite{scan:tcuICS19} was to improve the performance of the state-of-the-art GPU scan kernels (a very challenging task), whereas our focus is currently only limited to algorithmic design and analysis. Moreover, the authors mentioned that their approach works best for small segment sizes~\cite{scan:tcuICS19}, whereas our approach might scale to larger input sizes. Nevertheless, we attempt to compare Algorithm~\ref{alg:tcu_scan} against Algorithm~7 of~\cite{scan:tcuICS19} below. In addition, we assume that GEMM operations are considered basic operation in the analysis below.
Algorithm~7~of~\cite{scan:tcuICS19} is expressed for the particular case $s=16$, and it is assumed that each warp takes $256$ elements, and each block has at most $16$ warps. For comparison, we replace in~\cite[Alg.~7]{scan:tcuICS19}, the constants $16$ and $256$ with $s$ and $s^2$, respectively. If $n$ is large enough, Algorithm~7~of~\cite{scan:tcuICS19} serializes the processing on the block level (for loop in Lines 7-23 of~\cite[Alg.~7]{scan:tcuICS19}) and, hence, the depth of the algorithm is at least $5\lceil n/s^3\rceil$ since at most $s$ warps exist per block and $3$ GEMMS, an exclusive scan of size $16$ and a broadcast is required. Regarding work, each warp gets $s^2$ elements, and each GEMM operation requires $\OO(s^3)$ binary operations. Hence, the number of binary operations is $\OO(ns)$.
From an implementation point of view, the authors of~\cite{scan:tcuICS19} demonstrated that by taking advantage of the additional computational power offered by the tensor core units, it is possible to improve the performance of the state-of-the-art and high-performance scan implementations for small segment sizes on GPUs. On the other hand, we haven't yet developed a high-performance implementation of Algorithm~\ref{alg:tcu_scan}, but we plan to investigate such a direction in the near future. 
Table~\ref{tab:algo_comparison} summarises several representative parallel scan algorithms from the literature. Prior work is evaluated on the PRAM model, whereas Algorithm~\ref{alg:tcu_scan} and~\cite[Alg.~7]{scan:tcuICS19} are evaluated on the TCU model where we assume that multiplication of square matrices of constant size $s$ is a basic operation. As it is depicted in the table, for $s=4$, Algorithm~\ref{alg:tcu_scan} has depth $\log_2(n) -1$ in the TCU model. In this case, the work is $3n+\OO(\log_2(n))$ when simulated in the PRAM model (to have a fair comparison in terms of work with prior work). It is not possible to make a fair comparison in terms of depth in the PRAM model since the fan-in/fan-out of Algorithm~\ref{alg:tcu_scan} is also increased from two to four. Algorithm~7 of~\cite{scan:tcuICS19} has linear depth in the TCU model, and its work is $\OO(ns)$ when simulated in the PRAM model.
%

%
\section{Conclusion \& Future Work}\label{sec:conclusion}
%
We presented a parallel scan algorithm (\textsc{MatMulScan}) designed for the TCU model where matrix multiplication of square matrices of constant size is assumed to be a basic operation. A future research direction is to enlarge the applicability of the tensor core units to additional applications. Last but not least, we plan to design and develop a high-performant implementation based on \textsc{MatMulScan} using the Tensor Iterator Kernel (TIK) programming framework of the Ascend cube unit~\cite{huawei_ascend:hpca2021}.
%
%
%
%
%
\bibliographystyle{splncs04}
\bibliography{main}
%
%
\appendix
\section{Appendix}\label{appendix:tcu_scan_numpy}
%
We provide a functional end-to-end (but not high-performance) Python implementation of Algorithm~\ref{alg:tcu_scan} using NumPy (v1.24.1)~\cite{numpy:harris2020array}. The implementation demonstrates the memory layout operations required to orchestrate the batched matrix multiplications of Algorithm~\ref{alg:tcu_scan}.
%
\begin{lstlisting}[caption={Reference implementation of Algorithm~\ref{alg:tcu_scan}},captionpos=b,basicstyle=\small,frame=tb,language=python]
import numpy as np

def matmul_scan(x, s, k):
    L_s = np.tril(np.ones(s))
    B_s = np.eye(s)
    B_s[:, 0] = 1
    
    for t in range(k):
        start, step = s ** t - 1, s ** t
        y = x[start::step]
        z = batch_matmuls(y, L_s)
        x[start::step] = z
    
    for t in range(k - 1, 0, -1):
        start, step = s ** t - 1, s ** (t - 1)
        y = x[start::step]
        z = batch_matmuls(y, B_s)
        x[start::step] = z

def batch_matmuls(y, A_s):
    m, s = len(y), A_s.shape[0]
    y = y.flatten()
    extra_pad = int((s ** 2) * np.ceil(m / s ** 2))
    y.resize(extra_pad)
    
    T = y.reshape((-1, s, s)).transpose((0, 2, 1))
    W = A_s @ T # batched matrix multiplication
    z = np.reshape(W, (-1, s ** 2), order='F').flatten()
    return z[:m]    
\end{lstlisting}
%
%
\subsection{Correctness of Algorithm~\ref{alg:tcu_scan}}\label{appendix:correctness}
%
In this section, we prove the correctness of Algorithm~\ref{alg:tcu_scan}. We reformulate Algorithm~\ref{alg:tcu_scan} using recursion as stated in Algorithm~\ref{alg:recursive_scan}. The recursive formulation will enable us to prove the correctness using strong induction. Indeed, we prove by induction that \textsc{MatMulScanRecursive} is correct for all inputs that are powers of $s$, given an arbitrary $s\geq 2$.
%
\begin{algorithm}[!ht]
\begin{algorithmic}[1]
\Procedure{MatMulScanRecursive}{$\x$, $s$} \Comment{$s\geq 2$}
\State $\z \gets \Call{BatchMatMul}{\x, \matL_s}$ \Comment{ $\Call{BatchMatMul}$ of Algorithm~\ref{alg:tcu_scan}}
\State \textbf{Output:} Return $\Call{Recurse}{\z, s}$
\EndProcedure

\Procedure{Recurse}{$\z$, $s$}
\State If $\text{len}(\z) \leq s $, return $\z$ \Comment{Termination criterion}
\State $\text{start} \gets s-1$
\State $\z[\text{start}::s] \gets \Call{BatchMatMul}{\z[\text{start}::s], \matL_s}$
\State $\z[\text{start}::s] \gets \Call{Recurse}{\z[\text{start}::s], s}$
\State $\z[\text{start}:] \gets \Call{BatchMatMul}{\z[\text{start}:], \matB_s}$
\State \textbf{Output:} Return $\z$
\EndProcedure
\end{algorithmic}
\caption{Parallel Matrix Multiplication Scan (Recursive version)}\label{alg:recursive_scan}
\end{algorithm}
%

%
In particular, it suffices to show that the \textsc{Recurse} method with input $\z$ and $s$ has the following precondition/postcondition relation: given the precondition that on all consecutive chunks of size $s$ of $\z$, i.e., $(0,1,\dots, s-1), (s,s+1,\dots , 2s-1), \dots$, the ``local'' prefix sums on each chunk is precomputed, \textsc{Recurse} returns the prefix sum of $\z$ (postcondition). Indeed, by the definition of \textsc{MatMulScanRecursive} in Line 2 the ``local'' prefix sums of size $s$ are computed and, in Line 3, the \textsc{Recurse} method is called with the precondition to be true.
%

%
\paragraph{Base case.}
%
For inputs of size less than $s$, the termination criterion of Line $6$ is met, therefore the postcondition follows directly from the precondition since the input size is less than $s$.
%
\paragraph{Inductive step.}
%
The inductive hypothesis is that \textsc{Recurse} is correct for input sizes strictly less than $n$. We will show that \textsc{Recurse} is correct for inputs of size $n$. Indeed, given an input $\z$ where all its ``local'' prefix sums are precomputed, we prove that \textsc{Recurse} with input $\z$ returns the prefix sum of $\z$. Now, Line $8$ computes the ``local'' prefix sums on the $s$-strided subvector $\x[\text{start}::]$. The prefix sum on $\x[\text{start}::s]$ is computed on Line $9$ by the inductive hypothesis. Then, Line $10$ broadcasts and add the correct prefix sum values of the $s$-strided subvector of $\z$ to the corresponding $s$ following indices of each subvector. Hence, the postcondition of \textsc{Recurse} holds.
%
\end{document}